\documentstyle[12pt]{article}
\begin{document}
\renewcommand{\thefootnote}{\fnsymbol{footnote}}
\title{
	$(p+2)$-form gauge fields for $p$-branes
through the action-at-a-distance
}

\author{
	TADASHI MIYAZAKI,
	CHI\'E OHZEKI\footnotemark[1] \\
	and
	MOTOWO YAMANOBE\footnotemark[2] \\
	{\it Department of Physics, Science University of Tokyo,} \\
	{\it 1-3, Kagurazaka, Shinjuku-ku, Tokyo 162, Japan}
}
\footnotetext[1]{E-mail: ohzeki@grad.ap.kagu.sut.ac.jp}
\footnotetext[2]{E-mail: yamanobe@grad.ap.kagu.sut.ac.jp}

\maketitle
\vspace{20mm}
\vspace{20mm}

\begin{quote}
\begin{center}
{\bf Abstract}
\end{center}
In the conventional $p$-brane theory, 
a gauge ($p+1$)-form field $\phi^{(p+1)}$ mediates the interaction 
between $p$-branes, which is gauge invariant in the sense 
$\phi^{(p+1)}\to \phi^{(p+1)}+d\Lambda^{(p)}$ with $\Lambda^{(p)}$, 
an arbitrary $p$-form and $d$, a boundary operator. 
We, on the contrary, propose to introduce a 
new gauge field $\phi^{(p+2)}$ mediating the 
interaction, and we have a new type of gauge
 transformation: $\phi^{(p+2)}\to \phi^{(p+2)}
+\delta\Lambda^{(p+3)}$, 
with $\delta$, a coboundary operator. 
\end{quote}

\newpage

Starting with the study of 
a one-dimensional object ({\it string}), 
one concerns oneself 
with an arbitrary-dimensional extended object 
({\it p-brane}). 
With the $p$-brane theory such an ambitious problem 
as unifying all fundamental interactions, 
which, as a result, clarifies 
the relation between $p$-brane interaction and supergravity, is now 
going to be attacked. 

We, in the present report, 
want to go along this line 
and to deal with the interaction between $p$-branes. 
The gauge field of rank ($p+1$) couples 
to $p$-branes, whose way of coupling is invariant 
under the conventional gauge transformation.
The interaction Lagrangian of $p$-branes 
and gauge fields, to say noting of 
the kinetic Lagrangian, is made to be invariant under 
the conventional gauge transformation 
$\phi^{(p+1)}\to\phi^{(p+1)}+d\Lambda^{(p)}$, with an arbitrary 
$p$-form $\Lambda^{(p)}$ ($d$ : a boundary operator). 
This interaction to the $p$-branes of an arbitrary dimension 
has been described in great detail 
in Refs.\cite{open}\cite{close}. 

However, the question arises as to 
whether there exists an 
{\it alternative interaction} or not 
which also describes the $p$-brane 
motion. The answer is affirmative 
and we will introduce a new type of 
gauge transformation 
$\phi^{(p+2)}\to\phi^{(p+2)}+\delta\Lambda^{(p+3)}$, with 
$\Lambda^{(p+3)}$, an arbitrary $(p+3)$-form, 
and $\delta$, a coboundary 
operator dual to $d$. 
The gauge transformation of this kind appears in 
the theory with magnetic monopoles\cite{miyazaki}. 

Feynman and Wheeler succeeded in formulating 
a theory equivalent to the Maxwell electromagnetic theory 
through the action-at-a-distance (AD) 
force\cite{feynman}. 
They obtained, with this AD force, the electromagnetic 
field (gauge field) mediating 
the charged particle interaction. 
Ishikawa {\it et al.} applied the method to 
the $p$-brane theory and 
wrote down the interaction Lagrangian 
with a $(p+1)$-form gauge field 
mediating the $p$-brane interaction\cite{open}\cite{close}. 
There the gauge transformation was defined 
{\it \'{a} la} Feynman and Wheeler, i.e., 
in the conventional form in electromagnetism. 
Now, in this report, 
we will formulate an essentially 
new type of gauge transformation. 

Following the reference\cite{open}, 
we consider two point particles $a$ and $b$, 
interacting with each other through 
the AD force in $D$-dimensional 
space-time with metric ($g^{\mu\nu}$)=diag(\\ 
$-$,$+$,$\cdots$,$+$). 
The action of this system is given by 
\begin{equation}
           S=S_{a}^{free}
             +S^{free}_{b}
             +\frac{g_{a}g_{b}}{2}\int 
             d\tau_{a}d\tau_{b}
             R_{ab}(X_{a},X_{b},
             \dot{X}_{a},\dot{X}_{b}),       \label{3.1.1}
\end{equation}
where $S_{a}^{free}$ (or $S_{b}^{free}$) 
is the action of 
the free {\it particle} $a$ (or $b$), 
$g_{a}$ (or $g_{b}$) is a coupling constant, 
and $\tau_{a}$ (or $\tau_{b}$) 
is a proper time of $a$ (or $b$) 
($\dot{X}_{a}\equiv 
dX_{a}/d\tau_{a}$, etc.). 
The existence of the functional $R_{ab}$ 
just represents that of the AD force. 
It satisfies $R_{ab}$=$R_{ba}$ 
for symmetry requirement. 
Imposing reparametrization invariance 
on the action, 
we adopt the following form:
\begin{eqnarray}
           R_{ab} &=& M_{a}^{\mu\nu}M_{b \mu\nu}
                    G((X_{a}-X_{b})^{2}),             \label{3.1.2}
\\
           M_{a\mu\nu} &\equiv& 
                        \dot{X}_{a\mu} \partial_{a\nu}
                        -\dot{X}_{a\nu} \partial_{a\mu},
\end{eqnarray}
where $G((X_{a}-X_{b})^{2})$ is 
a Green's function (with a mass parameter $m$) 
in the $D$-dimensional space-time. 
By assuming that the action (\ref {3.1.1}) 
be stationary
 under the variation
\begin{equation}
        X^{\mu}_{a}\to X^{\mu}_{a}
                  +\delta X^{\mu}_{a},                \label{3.1.3}
\end{equation}
we obtain the equations of motion for 
the particle $a$,
\begin{equation}
              \ddot{X}^\mu_a = g_a
              \partial_{a\rho}\left(
              \partial_{a}^{\mu}
              g_{b}\int d\tau_{b}M_b^{\nu\rho}
              G((X_a-X_{b})^{2})
              -\partial_{a}^{\nu}
              g_{b}\int d\tau_{b}M_b^{\mu\rho}
              G((X_a-X_{b})^{2})
              \right)\dot{X}_{a\nu}.               \label{3.1.3-2} \\
\end{equation}
Note that these equations 
of motion are expressed in the particle coordinates 
and their derivatives, 
{\it without gauge fields}. 
Following Feynman and Wheeler, 
we define the gauge fields: 
\begin{equation}
              \phi_{b\mu\nu}(X)
              \equiv g_{b}\int d\tau_{b}M_{b\mu\nu}
              G((X-X_{b})^{2}),                      \label{3.1.4}
\end{equation}
and then the equations of motion reduce to
\begin{equation}
              \ddot{X}^\mu_a 
              = g_aF^{\mu\nu}_b\!(X_a)
              \dot{X}_{a\nu},                        \label{3.1.5} \\
\end{equation}
with the field strength
\begin{equation}
              F_{b}^{\mu\nu}(X) 
              \equiv 
              \partial_{\rho}(\partial^{\mu}
              \phi_{b}^{\nu\rho}(X)
              -\partial^{\nu}
              \phi_{b}^{\mu\rho}(X)).                \label{3.1.6}
\end{equation}
In terms of differential forms, Eq.(\ref{3.1.6}) 
is rewritten as 
\begin{equation}
              F^{(2)}_{b}=d\delta\phi^{(2)}_{b}                \label{3.1.60}
\end{equation}
with self-evident definition of 
$\phi^{(2)}$ and $F^{(2)}$ 
(e.g., $F^{(2)}
=1/2F_{\mu\nu}dX^{\mu}\wedge dX^{\nu}$) 
as well as a boundary operator $d$ 
and a coboundary operator 
$\delta$.
Then we immediately find that 
the equations of motion (\ref{3.1.5}) are invariant
under the transformation of gauge 2-form field,
\begin{equation}
\phi^{(2)}\to
          \phi^{(2)}
          +\delta\Lambda^{(3)},                      \label{3.1.8}
\end{equation}
\begin{center}
$\Lambda^{(3)}$ : an arbitrary 3-form.
\end{center}
In components, Eq.(\ref{3.1.8}) 
is written as
\begin{equation}
\phi_{b\mu\nu}\to \phi_{b\mu\nu}
             +\partial^{\alpha}
             (\Lambda_{\alpha\mu\nu}
             -\Lambda_{\alpha\nu\mu}),               \label{3.1.7}
\end{equation}
\begin{center}
         $\Lambda_{\alpha\mu\nu}$ : 
         an arbitrary antisymmetric tensor of rank 3.
\end{center}
We note the conventional gauge transformation is
\begin{equation}
             A^{(1)}\to 
             A^{(1)}+d\Lambda^{(0)},             \label{2.1.14}
\end{equation}
\begin{center}
$\Lambda^{(0)}$ : an arbitrary 0-form.
\end{center}
Here a boundary operator $d$ 
is dual to a coboundary one $\delta$ in Eq.(\ref{3.1.8}). 
The total action is given, 
omitting the indices $a$ and $b$ 
indicating the particular particles, by
\begin{equation}
            S=-\int\!d\tau
            \left(-\dot{X}^\mu
            \dot{X}_\mu\right)^{\frac{1}{2}}+
            g\int d\tau \dot{X}^{\mu}\partial^{\nu}
            \phi_{\mu\nu}-
            \frac{1}{4}\int 
            d^D X F^{\mu\nu}F_{\mu\nu}         \label{2.1.14-1}
\end{equation}
The mass terms of gauge fields do not appear
because of gauge invariance.

The extension of the above method to $p$-branes is now 
straightforward. 
The action of interacting $p$-branes through 
the AD force is
\begin{equation}
         S=S_{a}^{free}
           +S^{free}_{b}
           +\frac{g_{a}g_{b}}{p+2}
           \int d^{p+1}\!\xi_{a}\,d^{p+1}\!\xi_{b}
           R_{ab}(X_{a},X_{b},X_{a,i},X_{b,i}).       \label{2.1.14-2}
\end{equation}
The notation here is self-evident; 
the difference 
(and {\it complication}) 
arises only from the fact that 
our $p$-brane traces 
the $(p+1)$-dimensional worldvolume 
in the $D$-dimensional target Minkowski space-time.
Adding a few more formulation to make doubly sure, we have
\begin{eqnarray}
	S^{\rm free}
	& = & -\int d^{p+1}\xi
        (-\sigma\cdot\sigma)^{\frac{1}{2}}.     \label{def:Sfree}
\end{eqnarray}
\begin{eqnarray*}
	\sigma\cdot\sigma
	& \equiv &
	\sigma^{\mu_{0}\mu_{1}\ldots\mu_{p}}
	\sigma_{\mu_{0}\mu_{1}\ldots\mu_{p}}, \\
	\sigma^{\mu_{0}\mu_{1}\ldots\mu_{p}}
	& \equiv &
	\frac{\partial(X^{\mu_0},X^{\mu_1},\ldots,X^{\mu_p})}%
	{\partial(\xi^0,\xi^1,\ldots,\xi^p)}, \\
	d^{p+1}\xi
	& \equiv &
	d\xi^{0}d\xi^{1}\ldots d\xi^{p}, \\
	X_{,i}^{\mu}
	& \equiv &
	\frac{\partial X^{\mu}}{\partial\xi^{i}},
\end{eqnarray*}
where $\xi^i$ ($i$=0,1,...,$p$) are parameters of the worldvolume. 
As $R_{ab}$ we adopt the following form 
which is also an extension of Eq.(\ref{3.1.2}):
\begin{equation}
           R_{ab} = 
           M_{a}^{\mu_{0}\mu_{1}\cdots\mu_{p+1}}
           M_{b\mu_{0}\mu_{1}\cdots\mu_{p+1}}
           G((X_{a}-X_{b})^{2}), \\                    \label{3.2.9}
\end{equation}
where
\begin{equation}
            M_{a\mu_{0}\mu_{1}\cdots\mu_{p+1}} 
            \equiv
            \sum_{i=0}^{p+1}(-1)^{(p+1)\cdot i}
            \sigma_{a\mu_{i}\mu_{i+1}
            \cdots\mu_{p+1}\mu_{0}\mu_{1}
            \cdots\mu_{i-2}}
            \partial_{a\mu_{i-1}}.                 \label{3.2.10}
\end{equation}
The gauge field for $p$-branes is
\begin{equation}
             \phi_{b\mu_{0}\mu_{1}\cdots\mu_{p+1}}(X)
             \equiv g_{b}\int
             d^{p+1}\xi_{b}
             M_{b\mu_{0}\mu_{1}\cdots\mu_{p+1}}
             G((X-X_{b})^{2}).
\label{3.2.11}
\end{equation}
The field strength is given by
\begin{eqnarray}
             F_{b}^{\mu_{0}\mu_{1}
             \cdots\mu_{p+1}}\!(X)
             \equiv\sum_{i=0}^{p+1}(-1)^{(p+1)\dot i}
             \partial_{\nu}\partial^{\mu_{i}}
             \phi_{b}^{\mu_{i+1}\cdots\mu_{p+1}\mu_{0}
             \cdots\mu_{i-1}\nu}.
\end{eqnarray}
Based on the variational principle 
we have the equations of motion 
\begin{eqnarray}
              (p+1)D_{a\,\mu_{1}\cdots\mu_{p}}
              \left[\frac
              {\sigma_a^{\mu_{0}\mu_{1}\cdots\mu_{p}}}
              {(-\sigma_{a}\cdot\sigma_{a})^{\frac{1}{2}}}\right]
              = g_{a}\sigma_{a\mu_{1}\cdots\mu_{p+1}}
              F_{b}^{\mu_{0}\mu_{1}\cdots\mu_{p+1}}.        \label{3.2.12}
\end{eqnarray}
where
\begin{eqnarray}
	D^{\mu_{1}\mu_{2}\ldots\mu_{p}}
	\equiv
	\sum_{i=0}^{p}
	K^{\mu_{1}\mu_{2}\ldots\mu_{p}}_i
	\frac{\partial}{\partial\xi^{i}},               \label{def:D}
\end{eqnarray}
and
\begin{eqnarray}
	K^{\mu_{1}\mu_{2}\ldots\mu_{p}}_{i}
	\equiv
	\frac{\partial\sigma^{\alpha\mu_{1}\mu_{2}\ldots\mu_{p}}}%
	{\partial X_{,i}^{\alpha}}.                               \label{def:K}
\end{eqnarray}
In the definition of $K_{i}$,
we do not sum over the index $\alpha$.

The equations of motion 
(\ref{3.2.12}) are invariant under the gauge
transformations for $\phi_{\mu_0...\mu_{p+1}}$;
\begin{eqnarray}
            \phi_{\mu_{0}\mu_{1}
            \cdots\mu_{p+1}}\to 
            \phi_{\mu_{0}\mu_{1}\cdots\mu_{p+1}}
            +\partial^{\mu_{p+2}}
            \Lambda_{\mu_{0}\mu_{1}
            \cdots\mu_{p+1}\mu_{p+2}},                \label{3.2.13} 
\end{eqnarray}
\begin{center}
            $\Lambda_{\mu_{0}\mu_{1}
            \cdots\mu_{p+1}\mu_{p+2}}$ : 
            an arbitrary tensor of rank ($p$+3),
\end{center}
which, expressed in the differential form, become
\begin{equation}
            \phi^{(p+2)}  
            \to  \phi^{(p+2)}
            +\delta\Lambda^{(p+3)},                  \label{3.2.14}
\end{equation}
\begin{center}
$\Lambda^{(p+3)}$ : an arbitrary ($p$+3)-form.
\end{center}

Now we finally have the total action 
for interacting $p$-branes
\begin{eqnarray}
          S=\!& - &\! \int d^{p+1}\xi 
            (-\sigma\cdot\sigma)^{\frac{1}{2}}
             +g\int d^{p+1}\xi 
             \sigma^{\mu_{0}\mu_{1}\cdots\mu_{p}}
             \partial^{\mu_{p+1}}
             \phi_{\mu_{0}\mu_{1}\cdots\mu_{p+1}} \nonumber \\
             \!& - &\! 
             \frac{1}{2(p+2)!}\int 
             d^D X F^{\mu_{0}\mu_{1}\cdots\mu_{p+1}}
             F_{\mu_{0}\mu_{1}\cdots\mu_{p+1}}.          \label{total-p}
\end{eqnarray}

Now we come to the conclusion. Contrary to the conventional gauge 
transformation with a boundary operator $d$, our transformation 
is represented in Eq.(\ref{3.2.14}) with a coboundary operator 
$\delta$. $R_{ab}$, being written in the AD force of Eq.(\ref{3.2.9}), 
we are to have a gauge invariant theory. In Ref.\cite{open}, 
for an open $p$-brane, it was emphasized that the primitive 
form of gauge field as in Eq.(\ref{3.2.11}) should be 
modified to recover the gauge invariance of the action. 
Based on this modification, an interesting result was obtained 
that the massiveness does not prevent the gauge invariance, 
which is a conspicuous difference from the case of a closed 
$p$-brane. Our model shows, however, that 
no modification is necessary 
and the gauge invariance of the action holds in the form 
Eq.(\ref{3.2.11}), both for a closed and open $p$-brane. 

Of course, our choice for $R_{ab}$ (Eq.(\ref{3.2.9})), 
is not unique. Our guiding principle to obtain Eq.(\ref{3.2.9}) 
is that $R_{ab}$ should have as simple a form as possible 
and that it be invariant under a new type of gauge 
transformation(\ref{3.2.14}) with a coboundary operator 
$\delta$.

\bigskip
\begin{center}
{\bf Acknowledgment}
\end{center}
Tow of the authors (C.O. and M.Y.) would like to thank 
Iwanami F\^{u}jukai for financial support.



\begin{thebibliography}{9}
\bibitem{feynman}
R.~P.~Feynman and J.~A.~Wheeler, 
Rev. Mod. Phys. {\bf 24}, 425 (1949).
\bibitem{open}
S.~Ishikawa, Y.~Iwama, T.~Miyazaki, 
K.~Yamamoto, M.~Yamanobe and R. Yoshida, 
Prog. Theor. Phys. {\bf 96}, 227 (1996).
\bibitem{close}
S.~Ishikawa, Y.~Iwama, T.~Miyazaki and M.~Yamanobe, 
Int. J. Mod. Phys. {\bf A10}, 4671 (1996).
\bibitem{miyazaki}
T. Miyazaki, 
Hep-th/9609186
\end{thebibliography}
\end{document}